# Hund flat band in a frustrated spinel oxide

Dongjin Oh[1,†], Alexander Hampel[2], Joshua P. Wakefield[1], Peter Moen[3,4], Steef Smit[3,4], Xiangyu Luo[1], Marta Zonno[5], Sergey Gorovikov[5], Mats Leandersson[6], Craig Polley[6], Asish K. Kundu[7], Anil Rajapitamahuni[7], Elio Vescovo[7], Chris Jozwiak[8], Aaron Bostwick[8], Eli Rotenberg[8], Masahiko Isobe[9], Manish Verma[10], Matteo Crispino[10], Martin Grundner[11,12], Fabian B. Kugler[2], Olivier Parcollet[2,13], Ulrich Schollwöck[11,12], Hidenori Takagi[9,14,15], Andrea Damascelli[3,4], Giorgio Sangiovanni[10], Joseph G. Checkelsky[1], Antoine Georges[2,16,17,18] and Riccardo Comin[1,†]

[1]Department of Physics, Massachusetts Institute of Technology, Cambridge, MA 02139, USA
[2]Center for Computational Quantum Physics, Flatiron Institute, New York, New York, 10010, USA
[3]Department of Physics & Astronomy, University of British Columbia, Vancouver, Canada BC V6T 1Z1
[4]Quantum Matter Institute, University of British Columbia, Vancouver, Canada BC V6T 1Z4
[5]Canadian Light Source, Inc., 44 Innovation Boulevard, Saskatoon, SK, S7N 2V3, Canada
[6]Max IV Laboratory, Lund University, Lund, Sweden
[7]National Synchrotron Light Source II, Brookhaven National Laboratory, Upton, NY, USA
[8]Advanced Light Source, Lawrence Berkeley National Laboratory, Berkeley, CA 94720, USA
[9]Max Planck Institute for Solid State Research, 70569 Stuttgart, Germany
[10]Institut für Theoretische Physik und Astrophysik and Würzburg-Dresden Cluster of Excellence ct.qmat, Universität Würzburg, 97074 Würzburg, Germany
[11]Department of Physics and Arnold Sommerfeld Center for Theoretical Physics (ASC), Ludwig-Maximilians-Universität München, D-80333 Munich, Germany
[12]Munich Center for Quantum Science and Technology (MCQST), D-80799 München, Germany
[13]Université Paris-Saclay, CNRS, CEA, Institut de Physique Théorique, 91191, Gif-sur-Yvette, France
[14]Department of Physics, University of Tokyo, 113-0033 Tokyo, Japan
[15]Institute for Functional Matter and Quantum Technologies, University of Stuttgart, 70569 Stuttgart, Germany
[16]Collège de France, Université PSL, 11 place Marcelin Berthelot, 75005 Paris, France
[17]CPHT, CNRS, Ecole Polytechnique, IP Paris, F-91128 Palaiseau, France
[18]DQMP, Université de Genève, 24 quai Ernest Ansermet, CH-1211 Genève, Switzerland

[†]Corresponding authors: djoh@mit.edu and rcomin@mit.edu

**Abstract**

**Electronic flat bands associated with quenched kinetic energy and heavy electron mass have attracted great interest for promoting strong electronic correlations and emergent phenomena such as high-temperature charge fractionalization[1] and superconductivity[2]. Intense experimental and theoretical research has been devoted to establishing the rich non-trivial metallic and heavy fermion phases intertwined with such localized electronic states[3]. Here, we investigate the transition metal oxide spinel $LiV_2O_4$, an enigmatic heavy fermion compound lacking localized *f* orbital states. We use angle-resolved photoemission spectroscopy and dynamical mean field theory to reveal a new kind of correlation-induced flat band with suppressed inter-atomic electron hopping arising from intra-atomic Hund coupling. The appearance of heavy quasiparticles is ascribed to a proximate orbital-selective Mott state characterized by fluctuating local moments as evidenced by complementary magnetotransport measurements. The spectroscopic fingerprints of long-lived quasiparticles and their disappearance with increasing temperature further support the emergence of a high-temperature 'bad' metal state observed in transport data. This work resolves a long-standing puzzle on the origin of heavy fermion behavior and unconventional transport in $LiV_2O_4$. Simultaneously, it opens a new path to achieving flat bands through electronic interactions in *d*-orbital systems with geometrical frustration, potentially enabling the realization of exotic phases of matter such as the fractionalized Fermi liquids[4].**

The motion of electrons in many solids can be largely understood in terms of their kinetic energy. For example, non-interacting electrons can easily hop between atomic sites (a process characterized by a hopping amplitude *t*), allowing them to delocalize across the solid. The greater the hopping amplitude, the greater the kinetic energy and resulting bandwidth. Such itinerant electrons are described by the free electron model with quadratic dependence of the single-particle energy *E* on crystal momentum *k* for small *k* (Fig. 1a)[5]. In contrast, the kinetic energy of the non-interacting electrons is quenched in a geometrically frustrated lattice. For example, in the case of the kagome lattice, electron wave functions at two neighboring atomic sites have opposite phases and undergo quantum destructive interference (Fig. 1b). As a result, electrons are 'trapped' within the hexagon and form localized electron packets[6–8]. This mechanism of geometrical hopping frustration gives rise to an electronic flat band and enhances the electron effective mass.

In addition to lattice-driven interference, electronic motion can also be frustrated by strong interaction effects. In $f$-orbital systems, itinerant electrons ($i$-electrons) strongly interact with the localized, heavy $f$-electron through antiferromagnetic exchange coupling, the so-called Kondo interaction[9], as shown in Fig. 1c. The hybridization between these light and heavy electrons leads to the formation of heavy quasiparticle states (Fig. 1c). The correlation-induced electron mass enhancement can in principle occur even in the absence of $f$-orbital electronic states. Coulomb repulsion ($U$) and Hund's coupling ($J_H$), the latter being a pure interorbital effect, significantly renormalize the bandwidth ($W$) and electron mass to reduce the effective kinetic energy via the formation of dressed electronic quasiparticles (Fig. 1d)[10–12]. These forms of correlation-induced electron localization and mass enhancement have been extensively studied in strongly correlated electron systems, such as cuprate[13,14], iron-based[15], nickelate[16–18], and ruthenate superconductors[19,20]. Here, we investigate the possibility of realizing flat band states through interaction effects.

The realization of correlation-induced flat bands is of significant interest in condensed matter physics. A flat band is associated with a large density of states resulting from the reduced band dispersion. When an electronic flat band lies at the Fermi level, it makes the electronic ground state unstable toward ordered phases such as unconventional superconductivity[21], quantum magnetism, integer[22] and fractional quantum anomalous Hall effect[23], and anomalous Landau levels[24]. Due to these unique characteristics of electronic flat bands, significant efforts are directed toward realizing flat bands not only in crystalline solids[3,7,8,25–28] but also in photonic lattices[29,30], phononic[31] and magnonic crystals[32], and metal-organic frameworks[33]. Therefore, exploring new ways to construct electronic flat bands and broadening the materials platform that can host such states is a timely challenge and opportunity for quantum material research.

Lithium vanadate ($LiV_2O_4$) is an ideal candidate to explore novel kinds of flat bands. This transition metal oxide compound crystallizes in the spinel structure and its low-energy electronic structure is governed by $d$-orbital electrons. $LiV_2O_4$ is the material that displays heavy Fermi liquid behavior with heaviest mass despite not containing $f$-electron states[34–36]. Indeed, this material possesses all the ingredients for supporting an electronic flat band. The V ions of $LiV_2O_4$ comprise a frustrated pyrochlore network (see Fig. 1e) that can host frustration-induced compact localized states and a corresponding three-dimensional flat band[6,27]. Moreover, it has been argued that magnetic frustration within the pyrochlore lattice can also be an essential factor in emerging heavy fermion behavior[35,37–42]. Within the theoretical framework, strong electron correlation effects mediated by the Coulomb, Hund, and Kondo interactions have been actively discussed[38–40,42–46]. The electronic specific heat of $LiV_2O_4$ with large Sommerfeld coefficient γ ~ 420 mJ/mol K$^2$ (blue circles in Fig. 1f) is the first important fingerprint of heavy electrons[34,35] (uncorrelated metal typically have γ < 20 mJ/mol K$^2$, and heavy fermion with f-electron have γ ≈ 100 ~ 1000 mJ/mol K$^2$). Remarkably, the value of the Kadowaki-Wood ratio (the ratio of the coefficient of the $T^2$ term in electrical resistivity and γ$^2$) is in line with that of *f*-electron heavy fermion systems (Fig. 1g)[47]. In addition, temperature-dependent electrical resistivity data indicate incoherent 'bad' metallic behavior at high temperatures, with a crossover to a coherent metallic state below $T^* \sim 25$ K (red curve in Fig. 1f)[35]. A Fermi liquid state eventually materializes below 2 K[48]. These thermodynamic and electrical properties imply the presence of heavy quasiparticles at low temperatures. However, the precise microscopic mechanisms responsible for generating such heavy bands have remained elusive.

Here, we present new experimental and theoretical findings that provide conclusive insight into the origin of heavy fermion behavior in LiV$_2$O$_4$. We investigate the electronic band structure of LiV$_2$O$_4$ in three-dimensional (3D) momentum space using temperature- and photon energy-dependent angle-resolved photoemission spectroscopy (ARPES) experiments. ARPES data provide direct evidence for a coherent electronic flat band at the Fermi level. We further find that this coherent heavy quasiparticle state partly survives in the bad metallic regime up to 140 K, becoming fully incoherent above this temperature. Our experimental observation of the correlated flat band is supported by first-principles calculations based on the dynamical mean field theory (DMFT) framework, which identify orbital differentiation driven by Hund's coupling $J_H$ as the crucial mechanism that drives the formation of the flat band and associated heavy fermion behavior in LiV$_2$O$_4$. The orbital-specific correlated flat band results in a highly localized electron state with fluctuating local moment as evidenced by negative magnetoresistance and in agreement with other experimental results on LiV$_2$O$_4$[35,37]. Our study not only reveals the origin of heavy fermion behavior in this specific compound but may provide a generalizable strategy to produce heavy electronic bands in *d*-electron systems. Furthermore, the direct observation of heavy quasiparticles in a frustrated lattice may serve as a foundation for the experimental realization of novel phases of matter such as a fractionalized Fermi liquid[4,49].

## Results

### Observation of a flat band at Fermi level

ARPES spectra of LiV$_2$O$_4$ are shown in Fig. 2. A clear Fermi surface near the Γ point ($k_z$ = 0 Å$^{-1}$) was observed with several Fermi pockets consistent with the metallic behavior (Fig. 2a).

To map out the three-dimensional electronic band structure along $k_x$, $k_y$, and $k_z$ crystal momenta across the 3D Brillouin zone (Fig. 2b), we performed photon energy-dependent ARPES measurements. Fig. 2c and d show the energy-momentum dispersion obtained at 6 K and 165 K, respectively, along the L-Γ-L $k$-path aligned to the $k_z$ axis in the experimental coordinate system (see Fig. 2b). Both the low- and high-temperature spectra exhibit a broad electron-like conduction band between –0.4 ~ 0 eV binding energy near the Γ point ($h\nu$ = 95 eV). The strong modulation of the spectral weight with respect to the photon energy ($k_z$ momentum) reflects the three-dimensional nature of the electronic band structure of $LiV_2O_4$. Interestingly, the low-temperature ARPES spectrum exhibits an additional sharp peak at the Fermi level over a wide $k$-path from $h\nu$ = 85 eV ($k_z$ = –0.43 π/c) to 110 eV ($k_z$ = 0.5 π/c) (46 % of the Brillouin zone size, along L-Γ-L). The energy distribution curves in Fig. 2e clearly display the emergence of a sharp quasiparticle peak at the Fermi level at low temperature, providing direct signature of an electronic flat band along the L-Γ-L high symmetry line. Furthermore, ARPES spectra aligned to the in-plane momentum $k_x$ confirm the presence of an extended flat band. Fig. 2f shows the measured in-plane ARPES spectra at various out-of-plane $k_z$ momenta, with the flat spectral weight between $k_z$ = –0.43 and 0.32 π/c marked by red arrows. These systematic ARPES measurements directly demonstrate the formation of a low-energy flat band in $LiV_2O_4$, consistent with the reported heavy fermion behavior.

**Temperature dependence of the flat band**

We further carried out temperature-dependent ARPES measurements to trace the evolution of the flat band. Fig. 3a-f show the in-plane energy-momentum dispersion near the Γ point measured from 6 K to 160 K. At 6 K, a sharp spectral weight of the flat band is discerned within $k_x \approx \pm 0.4$ Å$^{-1}$ (50 % of the high symmetry K-Γ-K $k$-path), on top of a broad hump. The presence of sharp spectral features at the Fermi level indicates the formation of coherent quasiparticles, as per Landau's Fermi liquid formalism. Therefore, the observed sharp flat band in the low-temperature ARPES data can be identified with heavy coherent quasiparticles in $LiV_2O_4$. The coherent spectral weight is significantly renormalized with temperature, as shown in the energy distribution curves (EDCs). The sharp coherent peak at the Fermi level gradually disappears with increasing temperature and becomes almost indiscernible around 160 K (Fig. 3g). The spectral weight of the flat band, defined as the intensity in the gray area of Fig. 3g subtracted by that of 160 K, shows a loss of coherence with increasing temperature (Fig. 3h). Furthermore, the width of the flat band along $k_x$ also becomes gradually narrower upon increasing temperature as shown in Fig. 3i (see Supplementary Fig. 3). We note that the coherent heavy quasiparticle band is robust even in the bad metallic regime ($T > T^* \sim 25$ K), which is counterintuitive to the notion that a coherent quasiparticle state cannot exist within a bad metallic regime, similar to the behavior of the resilient quasiparticle[50]. Our ARPES results suggest that understanding the flat band formation mechanism requires consideration of effects beyond the single-particle picture, as achieved by our DMFT analysis.

**Spectral function calculation**

To elucidate the microscopic origin of the correlated flat band in $LiV_2O_4$, we calculate the momentum-dependent spectral function via a combination of density functional theory plus dynamical mean field theory (DFT+DMFT), which is a reliable methodology to account for many-body electron correlation effects[51]. The realistic Coulomb interaction obtained from first-principle calculations and the DMFT impurity problem is solved by the numerically exact QMC solver. Fig. 4a shows the DMFT spectral function at $T = 11.6$ K, which displays a sharp, flat, and extended quasiparticle band very close to the Fermi level ($E = -1.75$ meV at $\Gamma$). In $LiV_2O_4$ (as well as other oxide spinels), V ions are surrounded by oxygens, forming $VO_6$ octahedra with trigonal distortion (Supplementary Fig. 4). This trigonal distortion splits the $t_{2g}$ manifolds into an $a_{1g}$ singlet and an $e_g^\pi$ doublet. The orbital projected spectral function calculation identifies the narrow flat band mainly composed of the $a_{1g}$ orbital while the $e_g^\pi$ orbital exhibits dispersive bands. The temperature-dependent orbital projected local spectral functions shown in Fig. 4b demonstrate the emergence of a quasiparticle peak in the $a_{1g}$ orbital manifold below 140 K, consistent with ARPES results (inset of Fig. 4b). Below 38 K, the $a_{1g}$ spectral function shows a rapidly growing heavy quasiparticle peak (HQP) at the Fermi level while the $e_g^\pi$ orbital bands undergo minor renormalization (see also Supplementary Fig. 5). We note that the correlated $a_{1g}$ flat band does not arise from the compact localized states of the frustrated pyrochlore lattice. At the DFT level, an electronic flat band derived from the compact localized state of $e_g^\pi$ orbital character is well above the Fermi level and is linked to the typical band structure of the pyrochlore lattice[52]. Also in DMFT, the flat band of the $e_g^\pi$ orbital remains 0.6 eV above the Fermi level (Supplementary Fig. 5). Therefore, the $a_{1g}$ flat band is a purely correlated electronic flat band, in contrast to that recently reported flat band in chalcogenide spinel $CuV_2S_4$ driven by geometrical hopping frustration[53]. It is important to note that while our DMFT spectral function reveals a correlated flat band spanning the entire momentum space, our ARPES spectra capture only a part of it. This might be due to the finite dispersion of the flat band. According to our DMFT calculation, the flat band exhibits weak electron-like dispersion near the Γ point within a narrow energy range near the Fermi level (Supplementary Fig. 6). Due to this dispersion, flat band is positioned below the Fermi level within the momentum range of $k_x = \pm0.5$ Å$^{-1}$. Indeed, our ARPES spectra obtained at base temperature clearly captured the flat band spectral weight within the range of $k_x$ = -0.41±0.05 Å$^{-1}$ to $k_x$ = +0.41±0.05 Å$^{-1}$. Similarly, the $k_z$ dispersion of the flat band also reveals electron-like dispersion. Thus, the flat band lies below the Fermi level at the Γ point and crosses the Fermi level near $k_z$ = ±0.71 π/c. Our ARPES spectral weight confirm the flat band dispersion within the momentum range of $k_z$ = –0.43 π/c to $k_x$ = 0.5 π/c. While the ARPES data captures the flat band spectral weight in the anticipated momentum ranges, minor mismatches are observed. These discrepancies are likely due to the photoemission matrix element effects.

**Discussion**

The DMFT results strongly indicate that orbital-dependent electron correlations are the most important ingredient in the formation of a correlated flat band in LiV$_2$O$_4$[52,54]. This orbital differentiation is a result of the combination of the unique crystalline symmetry of the oxide spinels and Hund's coupling $J_H$. In the non-interacting limit without considering electron correlation ($U$ and $J_H$), the a$_{1g}$ orbital has a relatively narrow bandwidth (~0.7 eV) and large peak near the Fermi level while the e$_g^\pi$ levels have a broad bandwidth (~1.9 eV) as shown in Fig. 4c. In this limit, the a$_{1g}$ and e$_g^\pi$ orbitals have 0.4 and 0.55 electrons per V ion, respectively[52]. When we consider $U$ (3.94 eV), the electron filling of the a$_{1g}$ orbital increases up to 0.7 electrons per V ion in contrast to the e$_g^\pi$ which is reduced to 0.4 electrons. The inclusion of $J_H$ (0.56 eV) further promotes the orbital differentiation by pushing more electrons into the a$_{1g}$ band, with an electron count of 0.9 electrons per V ion. Therefore, the action of $J_H$ is to make the a$_{1g}$ (e$_g^\pi$) states more localized (delocalized) and lowering kinetic energy. As a result, Hund's coupling brings LiV$_2$O$_4$ in proximity of an orbital selective Mott phase (OSMP)[11,55], with a hole doping of ~10%. However, due to the interorbital hybridization between a$_{1g}$ and e$_g^\pi$ bands, LiV$_2$O$_4$ does not enter the OSMP but rather exhibits the strongly correlated metallic state with heavy quasiparticles upon cooling[56,57].

This phenomenon is reminiscent of the general trend in Hund's metals where interorbital coupling is the dominant factor behind electron correlations[11,12,15,58]. The correlated metallic behavior of the a$_{1g}$ electrons is closely related to the Hund metal physics accompanied by Hund's coupling induced fluctuating local moment and paramagnetic (no long-range magnetic order) incoherent metallic behavior[59]. This fluctuating local moment of $LiV_2O_4$ is corroborated by the paramagnetic Curie-Weiss behavior and the nonsaturating incoherent transport at high temperatures[34,35]. We note that, in the spin fluctuation regime, an external magnetic field makes the transport more coherent by polarizing the disordered local moments and reducing scattering from spin fluctuations[60]. As a result, negative magnetoresistance (MR) is expected in the incoherent metal region. Magnetotransport measurements on $LiV_2O_4$ (Fig. 4d) confirm this scenario through the clear negative MR in the incoherent transport regime. The fluctuating local moment is supported by other experimental data, namely the presence of spin fluctuations in the incoherent metallic regime of the $LiV_2O_4$[37,42].

Fig. 4e summarizes the temperature evolution of the macroscopic phenomena occurring in $LiV_2O_4$. First, the heavy $a_{1g}$ orbital, which is highly localized by Hund's coupling $J_H$, forms a fluctuating local moment at high temperature and strongly suppresses the coherence stabilizing the incoherent bad metallic behavior[11,52,61]. As temperature decreases, a correlated $a_{1g}$ flat band associated with the heavy quasiparticle state occurs. When the local moment is completely screened at low temperatures, $LiV_2O_4$ displays heavy Fermi liquid behavior in electrical resistivity and Pauli behavior in magnetic susceptibility[34].

Through the combination of ARPES, magnetotransport, and DMFT, we conclusively demonstrate the presence of a correlation-induced flat band as the key ingredient in the heavy fermion behavior in $LiV_2O_4$. These observations highlight spinel oxides as promising material candidates to explore a new pathway towards correlated flat bands (Fig. 4f). As discussed above, the trigonal splitting of the octahedral crystal field in the spinel structure initiates the orbital differentiation. Then, $J_H$ boosts the crystal field splitting and orbital-dependent electron correlation[11]. In this respect, oxide spinels have an advantage over chalcogenide spinels due to their strong ligand field originating from the large electronegativity of oxygen. Therefore, noticeable orbital differentiation in $a_{1g}$ and $e_g^\pi$ orbitals can arise in oxide spinels in conjunction with $J_H$. When two orbitals satisfy the appropriate electron filling, correlated electronic states with heavy effective mass enhanced by a spin blocking mechanism and fluctuating local moment can be stabilized[52,59]. In this circumstance, bad-metallic behavior may appear as a by-product. This orbital-dependent correlation in $LiV_2O_4$, promoted by $J_H$, stands in stark contrast to the electronic correlation in the chalcogen spinel compound $CuV_2S_4$ where the crystal field splitting of $t_{2g}$ manifold is absent and the distinct electronic correlation effects between these two compounds may explain their contrasting material properties, such as absence of heavy fermion behavior in $CuV_2S_4$[53].

However, these requisites alone are not sufficient to make $d$-electrons sufficiently heavy. Previous studies have reported experimental evidence of heavy fermion behavior in $d$-orbital systems such as $Ca_{1.5}Sr_{0.5}RuO4$[62], $YMn_2Zn_{20}$[63], $(Y_{0.97}Sc_{0.03})Mn_2$[64], $Fe_3GeTe_2$[65], $FeTe$[66], $Cs(Fe_{0.97}Cr_{0.03})_2As_2$[57]. Interestingly, a trend can be identified whereby a reasonably large Sommerfeld coefficient appears only in materials without long-range magnetic order: FeTe and $Fe_3GeTe_2$, which have antiferromagnetism and ferromagnetism, respectively, exhibit the smallest Sommerfeld coefficient among the above series. This tendency implies that long-range magnetic order competes with $d$-orbital heavy fermion behavior, an effect that was systematically studied in Cr-doped $CsFe_2As_2$[57]. In that sense, the suppression of long-range magnetic order and the effective superexchange via geometrical spin frustration in the pyrochlore sublattice of oxide spinels may be an important factor in fostering $d$-orbital heavy fermion behavior[52]. Although most of the $AB_2O_4$ oxide spinels which have integer valence state in the $B$ site have been known as a Mott insulator, the heavy electronic state may be realized via systematic chemical composition engineering by inducing a mixed valence state with $n_V >$ 1 as in the case of $LiV_2O_4$. For example, systematic studies on the strongly correlated metal-to-insulator transition upon chemical substitution from $AV_2O_4$ ($A$ = Mg, Zn) with $n_V = 2$ to Li end ($n_V = 1.5$)[67,68] or electrochemical Li (de)intercalation to the $LiV_2O_4$ ($1 \leq n_V \leq 2$)[69,70] may enable further control and engineering of correlation-induced flat bands.

Recent discussions have highlighted the intimate connection between the flat band and Hund's coupling $J_H$ in the high-temperature superconductor $La_3Ni_2O_7$[18]. DMFT calculations reveal that $J_H$ promotes the formation of flat bands through orbital-dependent correlations. However, this DMFT on nickelate superconductor also predicts strong antiferromagnetic coupling of the local moment, which contrasts with $LiV_2O_4$, where weak spin correlations arise due to geometrical frustration. Similarly, another DMFT study on $CsCr_3Sb_5$ and $CrB_2$ also underscores the crucial role of $J_H$ in the heavy fermion behavior associated with d-orbital electrons[71]. While these theoretical works support the applicability of our proposed approach for implementing a heavy quasiparticle electronic state at the Fermi level in other systems, this theoretical proposal remains experimentally unverified in these materials[17,72].

Furthermore, there has been active discussion about how to screen the high-spin electrons in Hund's metals and enter the Fermi liquid state within the DMFT framework[73,74,58,59,75,76]. The frustrated oxide spinel compounds could serve as ideal material candidates for experimentally validating these novel theoretical models. Moreover, it was pointed out that exotic heavy Fermi liquid phase with fractionalized excitation favors nonmagnetic frustrated lattice systems[4].

Therefore, future research on $LiV_2O_4$ and other frustrated spinel oxides, aiming to realize a heavy fermion state via local correlation effects, may pave the way to discovering new quantum phases of matter.

**Methods**

**Single crystal growth and characterization**

Single crystals of $LiV_2O_4$ were grown following a previous report[77] out of a melt containing 58% by weight $LiV_2O_4$ powder and 42% by weight $Li_3VO_4$. The powders were placed in a platinum crucible and heated in a quartz tube under vacuum to 1050° C, held at this temperature for 1 day, and then cooled over 5 days to 930° C after which the furnace was shut off. $LiV_2O_4$ crystals were separated from the flux by washing with water. $LiV_2O_4$ powder was grown out of a reaction containing $Li_3VO_4$, $V_2O_3$, and $V_2O_5$ in a 4:9:1 molar ratio. The reactants were pelletized and wrapped in Au foil before being sealed under vacuum in a quartz tube and heated at 700° C for 2 days. $Li_3VO_4$ was obtained by a reaction between $Li_2CO_3$ and $V_2O_5$ in a 3:1 ratio in air at 800° C for 20 h. Single crystals used for the magnetotransport measurements as well at ARPES experiments carried out at the National Synchrotron Light Source II (NSLS II) were further annealed at 700° C in vacuum wrapped in Au foil also containing powder $LiV_2O_4$ for up to a week.

Magnetotransport measurements were taken in a 4 contact geometry with current in the (111) plane and field perpendicular to the (111) plane. To make ohmic contact to the samples, first a pattern of 3 nm Ti followed by 40 nm Au was deposited onto the sample surface which was then contacted with Au wired by silver paint. Data was collected with a standard ac lock-in method at a typical current of 2 mA at fields between –9 T and 9 T. All field-dependent data is symmetrized with respect to field direction.

Specific heat data was taken down to 1.8 K in a commercial cryostat using the thermal relaxation method.

**Angle-resolved photoemission spectroscopy (ARPES)**

ARPES experiments were performed at Beamline7.0.2 (MAESTRO) of the Advanced Light Source (ALS), the QMSC (Quantum Materials Spectroscopy Centre) beamline of the Canadian Light Source (CLS), the Bloch beamline of the MAX IV laboratory, and the ESM beamline of the National Synchrotron Light Source II (NSLS II). All ARPES spectra were obtained from the (111) surface with linear horizontal polarization (p-polarization). Due to the matrix element effect, $a_{1g}$ and $e_g^\pi$ bands are mainly observed in the first and second Brillouin zone, respectively (Supplementary Fig. 2). The LiV$_2$O$_4$ crystals were cleaved inside ultra-high-vacuum (UHV) ARPES chambers (~4×10$^{-11}$ torr). The obtained cleaved surfaces were always aligned to the [111] crystallographic axis. To minimize beam induced damage that occurs when vacuum ultra-violet (VUV) light irradiates the LiV$_2$O$_4$ crystals, the undulator of the synchrotron facility was intentionally detuned to reduce photon flux as needed. The energy and momentum resolutions were approximately 30 meV and 0.01 Å$^{-1}$, respectively. To estimate the value of $k_z$ momentum, an inner potential $V_0$ of 15 eV was used.

**Density functional theory + dynamical mean field theory (DFT+DMFT)**

DFT calculations are performed using the Quantum Espresso (QE) software package using the ONCVPSP norm-conserving pseudo-potentials[78,79] in conjunction with the Perdew-Burke-Ernzerhof (PBE) exchange-correlation functional. Structural parameters are fixed to the experimental values reported at 12 K $a_0$ = 8.22694(3) Å and $x_0$ = 0.26109(2)[34]. We use a wavefunction cut-off of 90 Ry, a density cut-off of 360 Ry, and a mesh of 11 × 11 × 11 $k$-points resulting in an energy error of < 1 meV per formula unit.

Further, we construct maximally localized Wannier functions using Wannier90[80] for the low-energy states around the Fermi level with dominantly V $d$-character. This allows us to define a low-energy Hamiltonian $H^{W90}(R)$ in real space, capturing the essential physics of the system realistically. Furthermore, we rotate the the Hamiltonian on each V-site into the crystal field basis to diagonalize both the local non-interacting Hamiltonian and the hybridization function. For high accuracy low-temperature calculations, we leverage Wannier interpolation, Fourier transforming $H^{W90}(R)$ to a dense 41 × 41 × 41 $k$-point mesh to avoid any $k$-discretization error.

To determine the effective Coulomb interaction for this low energy model we utilize the constrained random phase approximation[81] as implemented in the RESPACK code . We limit ourselves here to the static, ω = 0 limit and fit the obtained four-index Coulomb tensor to the symmetrized Kanamori form (including spin-flip and pair-hopping terms) with three independent parameters. The optimal fit yields $U$ = 3.94 eV, $U'$ = 2.83 eV, and $J$ = 0.56 eV.

To cover all $T$ regimes we use a combination of impurity solvers to solve the DMFT equations. For calculations mainly at high temperature down to 11.6 K (1/1000 eV), we use the continuous-time hybridization-expansion QMC solver as implemented in the TRIQS software library[83,84] and its interface to electronic structure codes TRIQS/DFTTools and

TRIQS/SOLID_DMFT[52,85,86]. High-frequency noise in the impurity self-energy is suppressed by fitting the tail of $\Sigma_{imp}$ up to fourth order in a window with moderate noise. The first two moments of the tail expansion, are measured to high precision directly in TRIQS/CTHYB[87]. All QMC calculations include all pair-hopping and spin-flip terms of the interaction. Second, the $T = 0$ TN-based solver from Refs.[88–90] was extended to $T > 0$ using thermal-state purification[91,92]. The TN solver is used down to 2.9 K (1/4000 eV), while keeping as in QMC all terms of the interaction. Third, to access the sub-1 K regime, we use NRG, similarly as in Refs. [93,94]. To make non-degenerate three-orbital NRG calculations tractable, we increase the local symmetry by neglecting the pair-hopping part of the interaction when using this solver[93]. To calculate the real frequency spectral functions shown in the main text we analytically continue the impurity self-energy using Padé approximants and calculate the lattice Green function on the real frequency axis. More technical details on the DMFT calculations and for each solver can be found in Ref. [52].

**Data and Code availability**

The data in the manuscript are available from the corresponding author upon reasonable request.

**Acknowledgement**

We appreciate fruitful discussions with S. Todadri, and Y. Kim. This work was supported, in part, by the Air Force Office of Scientific Research (AFOSR) under grant FA9550-22-1-0432 (ARPES, material synthesis and characterization), the Gordon and Betty Moore Foundation EPiQS Initiative (grant no. GBMF9070 to J.G.C.) (instrumentation development), and the NSF (DMR-2104964) (thermodynamic measurements). This research used resources of the Advanced Light Source, which is a DOE office of Science User Facility under contract no. DE-AC02-05CH11231. This research used resources from the ESM beamline of the National Synchrotron Light Source II, a U.S. Department of Energy (DOE) Office of Science User Facility operated for the DOE Office of Science by Brookhaven National Laboratory under Contract No. DE-SC0012704. We acknowledge MAX IV laboratory for time on Bloch beamline under proposal 20230565. Research conducted at MAX IV, a Swedish national user facility, is supported by the Swedish Research council under contract 2018-07152, the Swedish Governmental Agency for Innovation Systems under contract 2018-04969, and Formas under contract 2019-02496. This research was undertaken thanks in part to funding from the Max Planck–UBC–UTokyo Centre for Quantum Materials, and the Canada First Research

Excellence Fund, Quantum Materials and Future Technologies Program. This project is also funded by the Natural Sciences and Engineering Research Council of Canada (NSERC), the Canada Foundation for Innovation (CFI); the British Columbia Knowledge Development Fund (BCKDF); the Department of National Defence (DND); the Mitacs Accelerate Program; the QuantEmX Program of the Institute for Complex Adaptive Matter; the Canada Research Chair Program (A.D.), and the CIFAR Quantum Materials Program (A.D.). Use of the Canadian Light Source (Quantum Materials Spectroscopy Centre), a national research facility of the University of Saskatchewan, is supported by CFI, NSERC, the National Research Council (NRC), the Canadian Institutes of Health Research (CIHR), the Government of Saskatchewan, and the University of Saskatchewan. The DMFT work was funded in part by the Deutsche Forschungsgemeinschaft (DFG, German Research Foundation) under Germany's Excellence Strategy—EXC-2111—390814868. G. S., M. C. and M. V. acknowledge support from the Deutsche Forschungsgemeinschaft (DFG, German Science Foundation) through FOR 5249 QUAST (Project-ID 449872909), EXC2147 ct.qmat (Project-ID 390858490) and SFB 1170 ToCoTronics (Project-ID 258499086), respectively. P. M. acknowledges the support from NSERC; S.S. acknowledges support by the Netherlands Organisation for Scientific Research (NWO 019.223EN.014, Rubicon 2022-3). The Flatiron Institute is a division of the Simons Foundation.

**Author contributions**

D.O., A.H., J.W., J.G.C., A.G., and R.C. conceived this project. D.O. and X.L. performed ARPES experiments at the Advanced Light Source, MAX IV, and the National Synchrotron Light Source II with support from C.J., A.B., E.R., M.L., C.P., A.K.K., A.R. and E.V.; D.O., P.M. and S.S. conducted ARPES experiments at the Canadian Light Source with support from M.Z., S.G. and A.D.; D.O. analyzed ARPES data; J.P.W. and M.I. synthesized and characterized the samples with support from J.G.C. and H.T.; A.H., M.G., F. B. K., O.P., U.S., A.G., M.C., M.V., and G.S. performed DFT and DMFT calculations; All authors contributed to writing the paper. D.O., G.S., J.G.C., A.G., and R.C. oversaw the project.

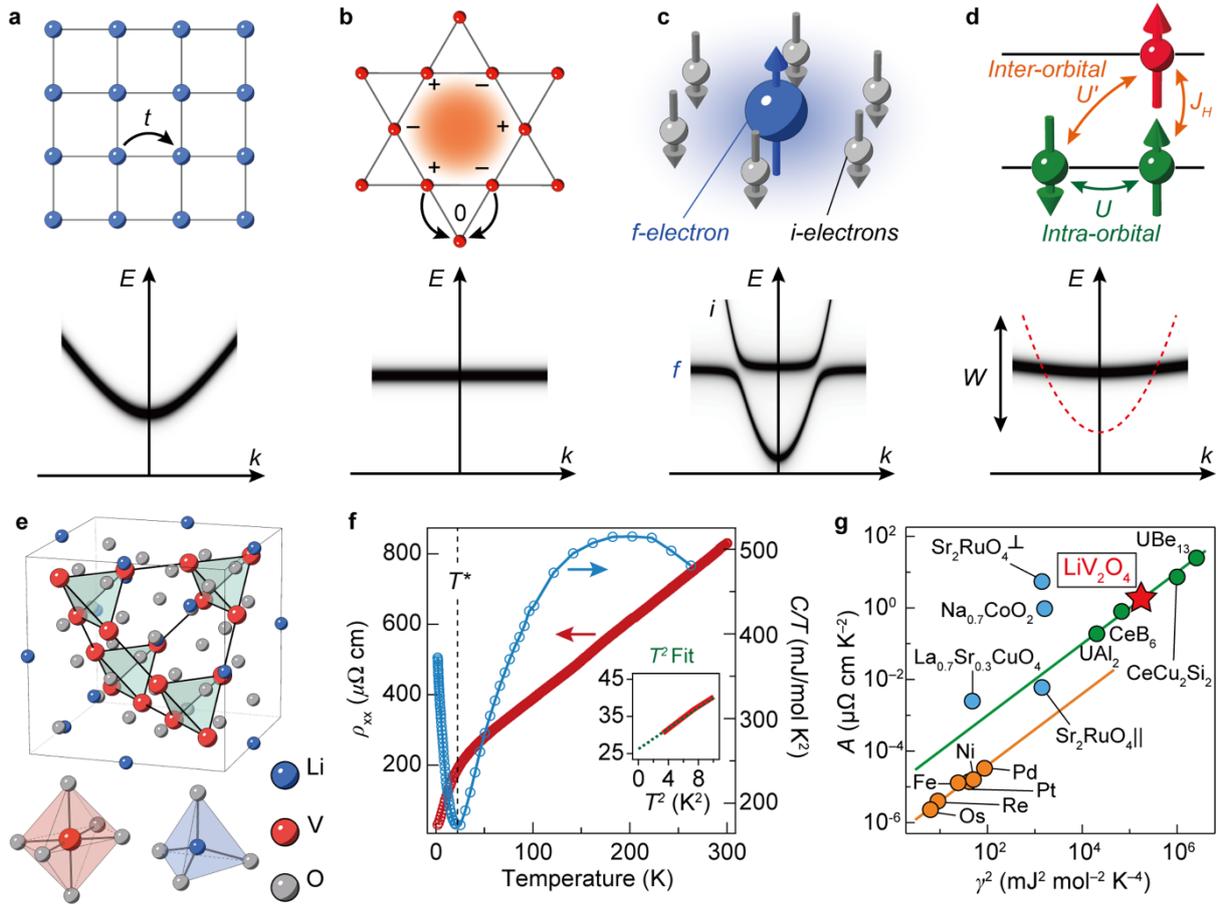

**Fig. 1| Heavy fermion behavior of LiV$_2$O$_4$. a-d,** Representative electronic band structures of the free electron model (a); flat band induced by geometrical frustration in the kagome lattice (b); hybridization between localized *f*-electron and itinerant electrons (*i*-electrons) (c); and bandwidth renormalization induced by strong electron correlation via intraorbital Coulomb ($U$), interorbital Coulomb ($U'$), and Hund's coupling ($J_H$) (d). The red dashed curve corresponds to the bare band dispersion without electron correlations. **e,** Crystal structure of LiV$_2$O$_4$ composed of VO$_6$ octahedra and LiO$_4$ tetrahedra. The V atoms form a frustrated pyrochlore network. **f,** Heavy Fermi liquid behavior of LiV$_2$O$_4$. The temperature-dependent electrical resistivity (red circles) shows a coherent-incoherent crossover while the electronic specific heat $C/T$ (blue circles) shows a large Sommerfeld coefficient $\gamma \approx 420$ mJ/mol K$^2$. Inset shows $\rho_{xx}$ versus $T^2$ plot (red) and $T^2$ fitting result (green). **g,** The Kadowaki-Wood ratio of various materials, defined as the ratio of the coefficient of the $T^2$ term in electrical resistivity and $\gamma^2$. LiV$_2$O$_4$ lies close to *f*-electron heavy fermion systems (green circles and line). Orange and blue circles represent the Kadowaki-Wood ratio of transition metals and oxide compounds, respectively. The value of the Kadowaki-Wood ratio was adapted from ref.47.

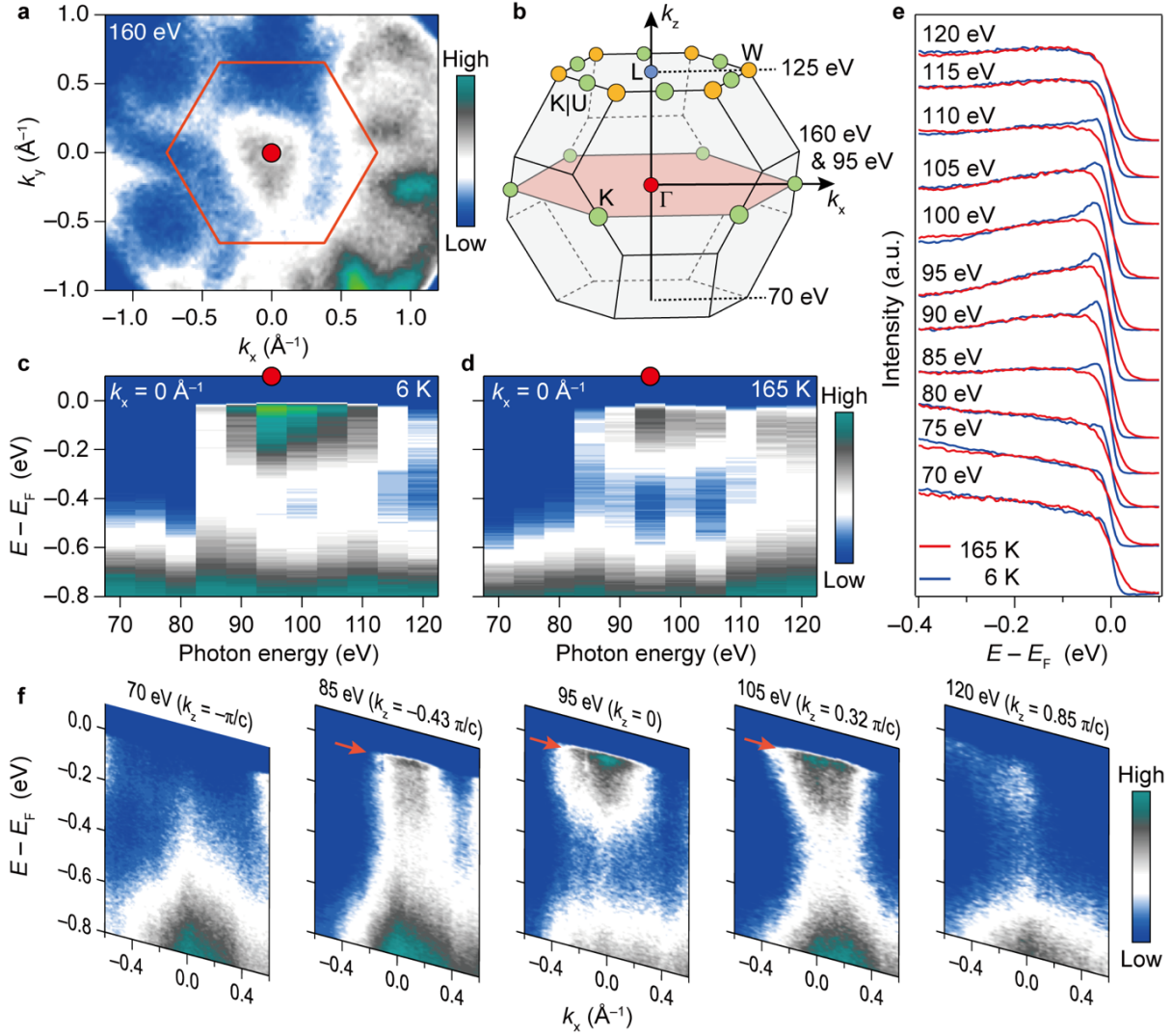

**Fig. 2| Observation of a flat band. a**, Fermi surface of LiV$_2$O$_4$ near the $k_z = 0$ Å$^{-1}$ plane. The red hexagon and circle represent the first Brillouin zone and Γ point, respectively. **b**, Three-dimensional bulk Brillouin zone. Red, green, blue, and yellow circles correspond to the high symmetry Γ, K|U, L, and W points, respectively. In the experimental geometry, the $k_x$ and $k_z$ momenta are aligned to the K-Γ-K and L-Γ-L directions, respectively. 95 eV and 160 eV photon energies probe the $k_z = 0$ Å$^{-1}$ momentum plane. On the other hand, 70 eV and 125 eV photon energies measure the $k_z = \pi/c$ plane. **c-d**, Low (c) and high (d) temperature energy-momentum dispersion across the L-Γ-L $k$-path aligned to the $k_z$ momentum. **e**, Comparison between low and high temperature energy distribution curves (EDCs) measured at the different photon energies. Red and blue EDCs were measured at 6 K and 165 K respectively. Each EDC is integrated within $k_x = \pm 0.1$ Å$^{-1}$. **f**, Photon energy-dependent energy-momentum dispersion along $k_x$, measured at 6 K. Red arrows indicate the observed flat band.

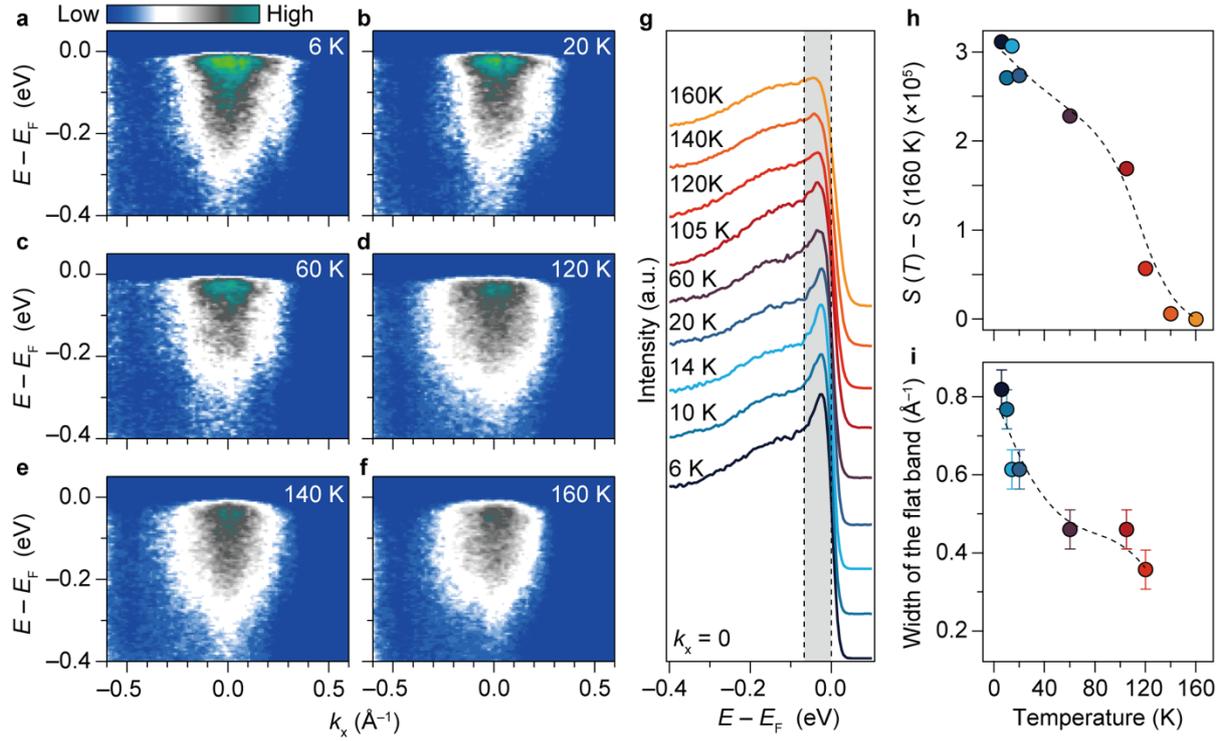

**Fig. 3| Temperature evolution of the flat band. a-f**, Temperature-dependent ARPES maps across the Γ point ($k_y$ and $k_z = 0$ Å$^{-1}$). **g**, Temperature-dependent energy distribution curves (EDCs) extracted from $k_x = 0$ Å$^{-1}$. Each EDC is integrated within $k_x = \pm 0.09$ Å$^{-1}$. The EDCs were normalized with the integrated intensities from −0.4 eV to −0.3 eV and shifted vertically. **h**, Temperature-dependence of the flat band spectral weight, defined as the integrated intensity in the gray area of (g) subtracted by the intensity at 160 K. **i**, Temperature-dependence of the width of the flat band, defined as the momentum range of the region where the quasiparticle peak is identified (see Supplementary Fig. 3). Error bars reflect the integrated momentum window.

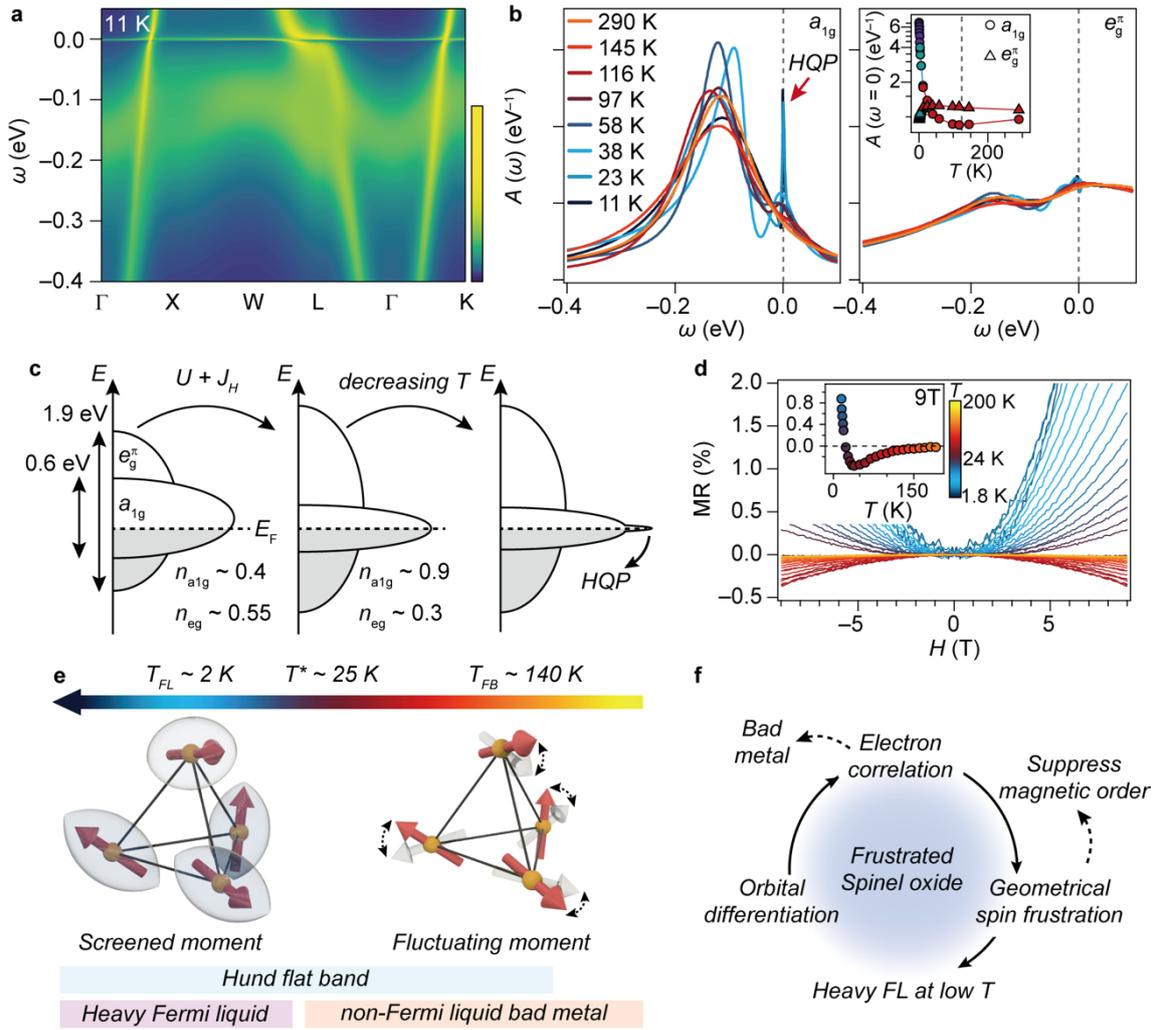

**Fig. 4| DMFT spectral functions and flat band formation mechanism. a**, Momentum-dependent spectral function of LiV$_2$O$_4$ across the high symmetry $k$-points. **b**, Momentum integrated spectral function of a$_{1g}$ (left) and e$_g^\pi$ (right) orbitals at various temperatures. The right panel inset represents the temperature-dependent intensity of spectral functions at the Fermi level. Circles and triangles indicate the intensity of a$_{1g}$ and e$_g^\pi$ orbitals, respectively. Purple, green, and red points are calculated by numerical renormalization group (NRG), tensor-network (TN), and quantum Monte Carlo (QMC) methods, respectively. **c**, Schematic energy diagram for the band structure renormalization. **d**, Magnetoresistance (MR), [$\rho_{xx}$ ($H$) − $\rho_{xx}$ (0)] / $\rho_{xx}$ (0), of LiV$_2$O$_4$ at various temperatures. Inset shows the temperature-dependent MR at 9 T for temperatures down to 16 K. **e**, Schematic temperature evolution of LiV$_2$O$_4$. Yellow spheres and red arrows represent V ions and a$_{1g}$ local moments. $T_{FL}$ and $T_{FB}$ are abbreviation for Fermi liquid and flat band onset temperature, respectively. **f**, Possible flat band formation mechanism in the frustrated spinel oxide.